\documentclass[runningheads]{llncs}
\usepackage[T1]{fontenc}
\usepackage{graphicx}
\graphicspath{ {./graphs/} }
\usepackage{subcaption}
\usepackage[inline]{enumitem}
\setlist[enumerate,1]{label={\roman*.}}
\usepackage{booktabs, multirow}
\usepackage{amsmath}
\DeclareMathOperator*{\argmin}{arg\,min}
\usepackage{hyperref, cleveref}
\usepackage{rotating}
\usepackage{placeins}
\usepackage{color}

\begin{document}
\title{Audio feature ranking for sound-based COVID-19 patient detection}%\thanks{Supported by organization x.}}
\author{Julia A. Meister\inst{1}\orcidID{0000-0003-2951-7217}
\and Khuong An Nguyen\inst{1}
\and Zhiyuan Luo\inst{2}}
\authorrunning{J.\ A.\ Meister et al.}
\institute{University of Brighton, East Sussex BN2 4GJ, United Kingdom \\
\email{\{J.Meister,K.A.Nguyen\}@brighton.ac.uk}
\and Royal Holloway University of London, Surrey TW20 0EX, United Kingdom \\
\email{\{Zhiyuan.Luo\}@rhul.ac.uk}}
\maketitle              % typeset the header of the contribution
\begin{abstract}
Audio classification using breath and cough samples has recently emerged as a low-cost, non-invasive, and accessible COVID-19 screening method. However, a comprehensive survey shows that no application has been approved for official use at the time of writing, due to the stringent reliability and accuracy requirements of the critical healthcare setting. To support the development of Machine Learning classification models, we performed an extensive comparative investigation and ranking of 15 audio features, including less well-known ones. The results were verified on two independent COVID-19 sound datasets. By using the identified top-performing features, we have increased COVID-19 classification accuracy by up to 17\% on the Cambridge dataset and up to 10\% on the Coswara dataset compared to the original baseline accuracies without our feature ranking.

\end{abstract}

\section{Introduction}\label{sec:intro}

    A widely accessible, non-invasive, low-cost testing mechanism is the number one priority to support test-and-trace in most pandemics. The advent of COVID-19 has abruptly brought respiratory audio classification into the spotlight as a viable alternative for mass pre-screening, needing only a smartphone to record a breath or cough sample~\cite{brown2020exploring}.
    
    It has long been common knowledge that respiratory diseases physically alter the respiratory environment in a way that often induces audible changes~\cite{rizal2015signal}. Consequently, manually listening to lung sounds (auscultation) is a common method to identify and diagnose respiratory disorders. However, many abnormalities only subtly affect auditory cues, making the inherently subjective auscultation process error-prone even when performed by a trained medical professional~\cite{aykanat2017classification}. To counteract subjectivity, automated audio classification approaches with promising results have become more and more common in recent years~\cite{amrulloh2015cough,aykanat2017classification,laguarta2020covid}.
    
    One of the main limiting factors is the lack of ground truth data which may be difficult to obtain, prone to limited population diversity, and requires medical training to label correctly. Because COVID-19 detection is a widespread and critical problem, multiple universities and research institutions have published COVID-19 audio datasets~\cite{brown2020exploring,sharma2020coswara}. This offers a unique opportunity to verify classification solutions on independently collected samples from a diverse population. The datasets have supported the development of a variety of applications with Machine Learning (ML) audio classification. However, at the time of writing, none have yet been officially endorsed for medical usage, largely because of the high accuracy and reliability expectations for such a critical healthcare task.
    
    The paper gives a comprehensive overview of relevant audio features (\Cref{sec:audioFeatures}) and identifies the most indicative ones for COVID-19 (\Cref{sec:experiments}). Finally, the findings are put into the context of existing literature (\Cref{sec:relatedWork}).

    \subsection{The paper's contributions}
        The rigorous feature analysis presented in this paper improves COVID-19 respiratory classification by optimising and holistically evaluating audio signal representations for Machine Learning (ML). The following contributions are made:

        \begin{itemize}
            \item \emph{Audio feature analysis and ranking}. We performed an extensive comparative analysis and ranking of 15 sound features prevalent and less-common in audio classification. The evaluation was carried out on two independent datasets, allowing the findings to be generalised.
            
            \item \emph{Highlighted effective features.} We identified sound-based ML features with strong discriminative performance that go against common rules of thumb.
            
            \item \emph{Increased COVID-19 detection accuracy.} We improved accuracy up to 17\% by incorporating new training features based on our feature ranking.
        \end{itemize}

\section{Audio features overview}\label{sec:audioFeatures}
    As in any Machine Learning (ML) application, feature engineering is a vital step for COVID-19 cough classification. We provide a detailed overview of 15 audio features from a variety of signal domains (\Cref{tab:featureSummary}) before rigorously evaluating their performance.

    \begin{table}[htbp]
        \small
        \centering
        \caption{\emph{Audio feature selection.} The 15 audio features evaluated in the paper.}\label{tab:featureSummary}
        \begin{tabular}{p{0.2\linewidth}p{0.245\linewidth}p{0.14\linewidth}p{0.38\linewidth}}
            \toprule
            \bfseries Domain & \bfseries Feature category & \bfseries  Name & \bfseries  Intuition \\
            \midrule
            \midrule
            \multirow{2}{*}{Time} & Signal energy & RMSE & Loudness of the signal. \\
             & Waveform & ZCR & Percussive vs\ tonal. \\
            \midrule
            \multirow{6}{*}{Frequency} & Spectral & S-BW & Perceived timbre. \\
             & Spectral & S-CENT & `Brightness' of a sound. \\
             & Spectral & S-CONT & Prevalence of formants. \\
             & Spectral & S-FLAT & Similarity to white noise. \\
             & Spectral & S-FLUX & Rate of frequency changes. \\
             & Spectral & S-ROLL & `Skewness' of the energy. \\
            \midrule
            \multirow{7}{*}{Time-frequency} & Cepstral & MFCC & Timbre, tone colour/quality. \\
             & Cepstral & MFCC-\(\Delta{}\) & Velocity of temporal change. \\
             & Cepstral & MFCC-\(\Delta{}^2\) & Acceleration of temporal change. \\
             & Tonal & C-ENS & Pitch. \\
             & Tonal & C-CQT & Pitch. \\
             & Tonal & C-STFT & Pitch. \\
             & Tonal & TN & Pitch \& pitch height. \\
            \bottomrule
        \end{tabular}
    \end{table}

    \subsection{Time domain}
        Low-level features extracted directly from the signal are in the time domain. They may identify crackling sounds caused by secretions in the throat and lungs~\cite{rizal2015signal}, and have been previously used for COVID-19 classification~\cite{brown2020exploring,sharma2020coswara}.

        \setlength{\abovedisplayskip}{3pt}
        \setlength{\belowdisplayskip}{3pt}
        \paragraph{Root mean square energy (RMSE).} A measure of the signal's amplitude over \(N\) frames, see Eq. (\ref{eq:rmse}). \(x_n\) is the average energy per frame~\cite{panagiotakis2005speech}.
                \begin{equation}\label{eq:rmse}
                    \textstyle
                    \textnormal{RMSE} = \sqrt{\frac{1}{N}\sum^{N}_{n=1}x^2_n}
                \end{equation}

        \paragraph{Zero-crossing rate (ZCR).} The signal's sign change rate (Eq. (\ref{eq:zcr})). \(x_n\) is amplitude at frame \(n\) of \(N\). \(sign(a)\) returns \(1\) if \(a > 0\), \(0\) if \(a = 0\), and \(-1\) else~\cite{panagiotakis2005speech}.
                    \begin{equation}\label{eq:zcr}
                        \textstyle
                        \textnormal{ZCR} = \frac{1}{2} \times \sum^{N}_{n=2}|sign(x_n)-sign(x_{n-1})|
                    \end{equation}

    \subsection{Frequency domain}
        To reveal frequency information of digital audio, it is decomposed into its constituent frequencies. This domain may identify abnormal lung sounds caused by an infection by examining the signal's intensity~\cite{rizal2015signal}. A subset has previously been used for COVID-19 detection~\cite{brown2020exploring,sharma2020coswara}.

        \emph{Spectral bandwidth.} Eq. (\ref{eq:sBw}) shows energy concentration, i.e.\ variance of expected frequency \(E\) given energy \(P_k\) and frequency \(f_k\) in \(1 \leq k \leq K\) bands~\cite{peeters2011timbre}.
                \begin{equation}\label{eq:sBw}
                    \textstyle
                    \textnormal{S-BW} = \sqrt{\sum^{K}_{k=1} (f_k - E^2 \times P_k)}
                \end{equation}
                
        \emph{Spectral centroid.} Eq. (\ref{eq:sCent}) shows the weighted and unweighted sums of spectral magnitudes \(P_k\) in the \(k\)-th of \(K\) subbands. \(f_k\) is the corresponding frequency~\cite{stolar2018detection}.
                \begin{equation}\label{eq:sCent}
                    \textstyle
                    \textnormal{S-CENT} = \frac{\sum^{K}_{k=1} P_k \times f_k}{\sum^{K}_{k=1}P_k}
                \end{equation}

        \emph{Spectral contrast.} Compare spectral peaks \(P_k\) and valleys \(V_k\) in frequency band \(k\), see Eq. (\ref{eq:sCont}). \(N\) is the number of frames and \(x'_{k,n}\) the FFT vector~\cite{jiang2002music}.
                \begin{equation}\label{eq:sCont}
                    \textstyle
                    \textnormal{S-CONT}_k = P_k - V_k = (\log \frac{1}{N} \sum^{N}_{n=1}x'_{k, n}) - (\log \frac{1}{N} \sum^{N}_{n=1}x'_{k, N-n+1})
                \end{equation}
                
        \emph{Spectral flatness.} Eq. (\ref{eq:sFlat}) measures similarity to white noise. \(P_k\) is the signal's energy at the \(k\)-th frequency band s.t. \(1 \leq{} k \leq{} K\)~\cite{madhu2009note}.
                \begin{equation}\label{eq:sFlat}
                    \textstyle
                    \textnormal{S-FLAT} = \frac{(\prod^{K}_{k=1}P_k)^{\frac{1}{K}}}{\frac{1}{K} \sum^{K}_{k=1} P_k}
                \end{equation}
                
        \emph{Spectral flux.} Eq. (\ref{eq:sFlux}) measures a signal's energy change between frames. \(E_{n,k}\) is the \(k\)-th of $K$ Discrete Fourier Transform coefficients in frame \(n\)~\cite{stolar2018detection}.
                \begin{equation}\label{eq:sFlux}
                    \textstyle
                    \textnormal{S-FLUX}_n = \sum^{K}_{k=1} E_{n,k} - E_{n-1,k}^2
                \end{equation}
                
        \emph{Spectral rolloff.} Eq. (\ref{eq:sRoll}) finds frequency \(f_R\) s.t.\ the energy accumulated below is no less than proportion \(S\) of total energy. \(P_k\) is energy in one of \(K\) bands~\cite{stolar2018detection}.
                \begin{equation}\label{eq:sRoll}
                    \textstyle
                    \textnormal{S-ROLL} = \argmin f_R \in \{1, \ldots, K\} \sum^{f_R}_{k=1} P_k \geq S \sum^{K}_{k=1} P_k
                \end{equation}

    \subsection{Time-frequency domain}
        This domain shows a signal's frequency as it varies over time. We consider two types of features: cepstral (timbre or tone colour) and tonal (pitch). 

        \textbf{Cepstral features.}  \enspace{}
            Non-linear Mel-frequency Cepstrum (MFC) is ubiquitous in respiratory classification because it explores a signal's temporal frequency content. It has been previously used for COVID-19~\cite{brown2020exploring,muguli2021dicova}.

            \emph{Mel-frequency cepstral coefficients.} Eq. (\ref{eq:mfcc}) shows the signal's transformation. \(s(k)\) is the log energy of the \(k\)-th of $K$ coefficients at frame \(n\)~\cite{brown2020exploring}.
                    \begin{equation}\label{eq:mfcc}
                        \textstyle
                        \textnormal{MFCC}_n = \sum^{K}_{k=1} s(k) \cos{\frac{\pi n (k-0.5)}{K}}
                    \end{equation}
                    
            \emph{MFCC-\(\Delta{}\).} The first-order derivative of MFCC, velocity, represents temporal change and is often included due to its low extraction cost~\cite{hossan2010novel}.
            
            \emph{MFCC-\(\Delta{}^2\).} The second-order derivative, acceleration, is commonly included because it may improve audio classification~\cite{hossan2010novel}.

        \textbf{Tonal features.}  \enspace{}
            Based on the human perception of periodic pitch~\cite{muller2005audio}. Two types are considered: chromagram and lattice graph. Secretions are a common consequence of COVID-19 which may alter the pitch of in- and expiration~\cite{rizal2015signal}.
            
            \emph{Chroma energy normalised.} Chroma abstraction considering short-time statistics within chroma bands. Normalisation makes C-ENS resistant to timbre~\cite{muller2005audio}.
            
            \emph{Constant-Q chromagram.} Extracted from a time-frequency representation. The constant-Q transform (C-CQT) has a good resolution of low frequencies~\cite{korzeniowski2016feature}.
            
            \emph{Short-time Fourier Transform chromagram.} The difference to C-CQT\@ is the initial transformation, in this case the Short-time Fourier Transform (STFT)~\cite{korzeniowski2016feature}.
            
            \emph{Tonnetz.} A lattice graph of harmonic information. Distances between points become meaningful by encoding pitch as geometric areas~\cite{humphrey2012learning}.

\section{Experimental method and results}\label{sec:experiments}

    The 15 investigated features range from prevalent to traditionally excluded from audio classification. They were ranked based on the empirical results analysis of two independent datasets. We assume that patterns repeated across both datasets are likely inherent to the COVID-19 respiratory recordings.

    \subsection{Research questions}\label{subsec:researchQuestions}
        Three research questions were formulated to inform the experimental design and results analysis. Each is focused on improving COVID-19 audio classification.

        \begin{itemize}
            \item What are the most predictive audio features for Machine Learning?
            \item Are the feature rankings comparable across independent datasets?
            \item How much does the performance accuracy of Machine Learning models improve by using the most dominant features?
        \end{itemize}

    \subsection{The datasets}\label{subsec:datasets}
        Two parallel independent datasets were considered throughout the paper to indicate whether feature rankings were likely generally applicable: the \emph{Cambridge} and \emph{Coswara} COVID-19 audio datasets. The sample counts are shown in \Cref{tab:sampleCounts}.
        
        Introduced in~\cite{brown2020exploring}, the \emph{Cambridge dataset} is a collection of healthy and COVID-positive cough and breath recordings. The data we used is a curated set of 48kHz WAV file samples, collected April-May 2020. Additionally, the Indian Institute of Science has collected shallow and deep breath and cough recordings in the \emph{Coswara dataset}~\cite{sharma2020coswara}. Compatible samples from April-December 2020 were considered. For consistency, we filtered for COVID-positive and healthy participants.

        \begin{table}[t]
            \small
            \centering
            \caption{\emph{Sample counts of the datasets.} Each Coswara participant has `shallow' and `deep' breath (B), cough (C), and breathcough (BC) recordings.}\label{tab:sampleCounts}
            \begin{tabular}{lrrrrrrrrr}
                \toprule
                \bfseries Label & \multicolumn{3}{c}{\bfseries Cambridge} & \multicolumn{3}{c}{\bfseries Coswara-deep} & \multicolumn{3}{c}{\bfseries Cos.-shallow} \\
                \cmidrule(lr){2-4} \cmidrule(lr){5-7} \cmidrule(l){8-10}
                & \multicolumn{1}{c}{B} & \multicolumn{1}{c}{C} & \multicolumn{1}{c}{BC} & \multicolumn{1}{c}{B} & \multicolumn{1}{c}{C} & \multicolumn{1}{c}{BC} & \multicolumn{1}{c}{B} & \multicolumn{1}{c}{C} & \multicolumn{1}{c}{BC} \\
                \midrule
                \midrule
                COVID-19 & 111 & 111 & 111 & 81 & 81 & 81 & 81 & 81 & 81 \\
                Healthy & 194 & 194 & 194 & 1074 & 1074 & 1074 & 1074 & 1074 & 1074 \\
                \(\sum{}\) & 305 & 305 & 305 & 1155 & 1155 & 1155 & 1155 & 1155 & 1155\\
                \bottomrule
            \end{tabular}
        \end{table}

    \subsection{Feature engineering}\label{subsec:featureEngineering}
        Cleaning the audio data was especially important because the recording devices and environments were not controlled. The pre-processing steps were carried out with the Python-toolkit \verb|librosa|, and included trimming the leading/trailing silences and normalising the amplitude to \((-1, 1)\).

\begin{figure}[!t]
            \centering
            \includegraphics[width=\textwidth]{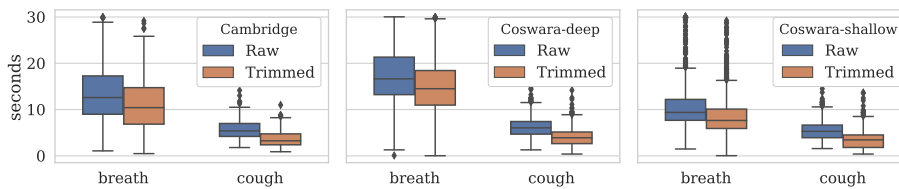}
            \caption{\emph{Sample lengths pre- and post-processing.} We trim leading and trailing silences (60dB, empirically identified). Lengths were reduced by 1--3 seconds.}\label{fig:sampleLengths}
        \end{figure}

        We evaluated 15 audio features from three signal domains (\Cref{sec:audioFeatures}). To standardise feature dimensions for Machine Learning (ML) models regardless of sample length (1--30 seconds, \Cref{fig:sampleLengths}), seven summary statistics were calculated to describe the feature distribution across frames: \begin{enumerate*}[label= (\roman*)] \item minimum, \item maximum, \item mean, \item median, \item variance, \item 1st quartile, and \item 3rd quartile\end{enumerate*}. Only a small subset of features was considered for evaluation and ranking at a time to avoid overfitting (812 features total, \Cref{tab:featureVector}).

        \begin{table}[!t]
            \centering
            \caption{\emph{Feature dimensions.} 812 features were considered. 7 Summary statistics  were taken across frames to ensure consistent dimensions (sample length 1--30s). To reduce overfitting risk, feature subsets were considered at a time for ranking.}\label{tab:featureVector}
            \begin{tabular}{cl}
                \toprule
                \bfseries Dimension & \bfseries Features (min, max, mean, median, var, Q\(_1\), and Q\(_3\))\\
                \midrule
                \midrule
                (1x7)  & RMSE, ZCR, S-BW, S-CENT, S-FLAT, S-FLUX, S-ROLL \\
                (6x7)  & TN \\
                (7x7)  & S-CONT \\
                (12x7) & C-ENS, C-CQT, C-STFT \\
                (20x7) & MFCC, MFCC-\(\Delta{}\), MFCC-\(\Delta{}^2\) \\
            \bottomrule
            \end{tabular}
        \end{table}

    \subsection{Results description and analysis}
        We identified the most informative features by evaluating two datasets in parallel. We propose that recurring predictive patterns are likely independent of the dataset, and should be strongly considered for future ML COVID-19 classification applications. Features were analysed in the following configurations:

        \begin{itemize}
            \item The Cambridge, Coswara-deep, and Coswara-shallow datasets.
            \item Breath (B), Cough (C), and BreathCough (BC) feature vectors. The latter is a concatenation of the previous two feature vectors, i.e.\ double the size.
            \item 5 models, selected for the variety in which they partition the label space: AdaBoost-Random Forest (ADA), K-Nearest Neighbours (KNN), Logistic Regression (LR), Random Forest (RF), and Support Vector Machine (SVM).
        \end{itemize}
        
        5-fold Cross-Validation ensured reliable results. We selected 3 metrics to compare the features' efficiency: \emph{Receiver Operating Characteristic} (ROC), \emph{Precision} (P), and \emph{Recall} (R). PR curves are well suited to imbalanced data by omitting true negatives, counteracting ROC's optimism~\cite{saito2015precision}. The mean over folds was a suitable indicator because the performance values passed the normality test \cite{iantovics2021black}.

\begin{figure}[p]
                \centering
                \begin{subfigure}[b]{\textwidth}
                    \centering
                    \includegraphics[width=\textwidth]{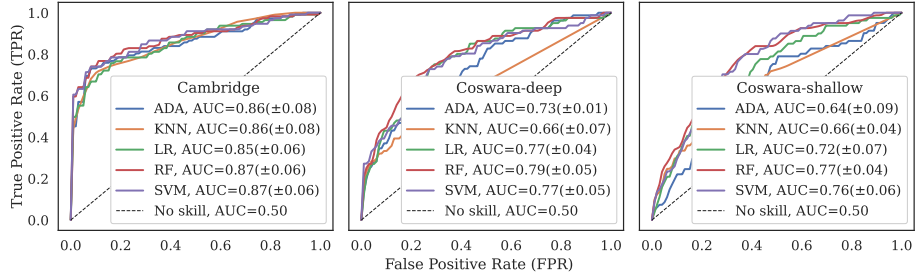}
                    \caption{Mean ROC over 5-fold CV (positive: COVID). AUC is `Area Under Curve'.}\label{subfig:breathcoughAllModelsRoc}
                \end{subfigure}
                \vfill %
                \begin{subfigure}[b]{\textwidth}
                    \centering
                    \includegraphics[width=\textwidth]{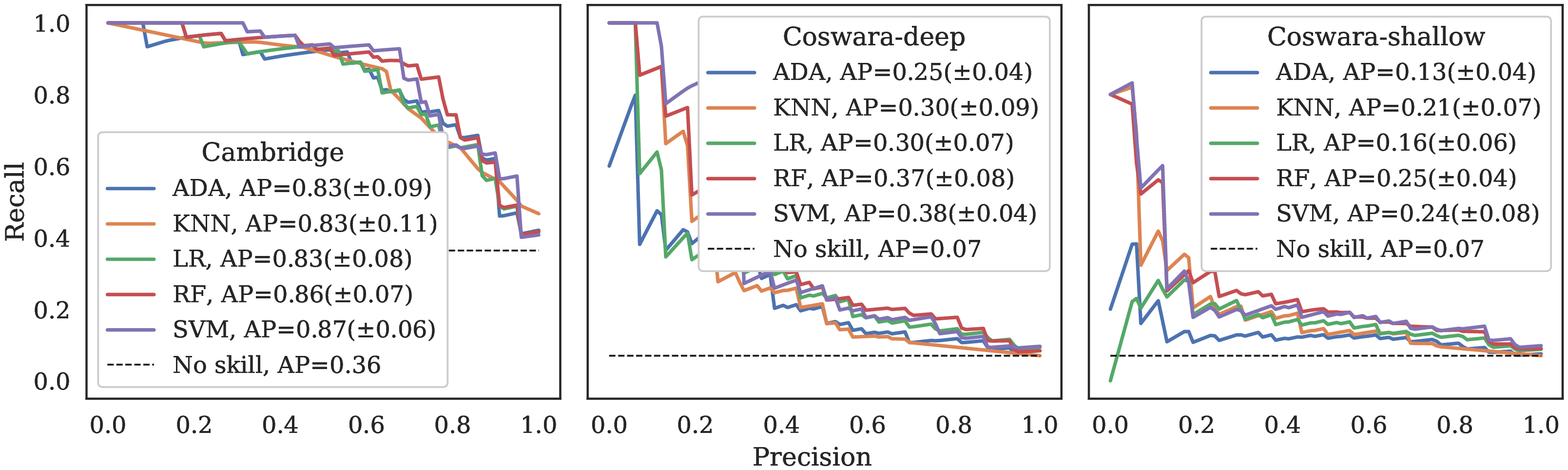}
                    \caption{Mean PR over 5-fold CV (positive: COVID). AP is `Average Precision'.}\label{subfig:breathcoughAllModelsPrc}
                \end{subfigure}
                \caption{\emph{BreathCough results.} Even though the ROC-curves look similar across datasets, the PR-curves reveal that Cambridge performed better overall. We also identified SVM and RF as the top-performing models. In PR-curves, the unskilled classifier corresponds to the dataset’s positive label ratio.}\label{fig:initalModelSelectionRocPcr}
            \end{figure}

        \textbf{Feature categories.} \enspace{}
            An overview of full feature vectors showed promising results, as most models outperformed their no-skill equivalent in ROC and PR-curves (\Cref{fig:initalModelSelectionRocPcr}). SVM and RF outperformed their counterparts across BC, B and C. Even though the two datasets had similarly shaped ROC curves, Cambridge had the best Average Precision (AP), and illustrates ROC's optimism on imbalanced datasets. An influential factor in Coswara's lower overall accuracies was the greater imbalance of COVID samples (Coswara 13:1 vs Cambridge 2:1, \Cref{tab:sampleCounts}). Nonetheless, Coswara-trained models performed significantly better than their unskilled classifier counterparts (13--38\% vs 7\% AP, \Cref{subfig:breathcoughAllModelsPrc}).

            \begin{table}[p]
                \small
                \centering
                \caption{\emph{BreathCough 5-fold CV ROC-AUC results as mean(std).} SVM and RF achieved the highest accuracies across most domains. The feature categories were be ranked in the following increasing order: time, tonal, spectral, cepstral.}\label{tab:featureCategoryResults}
                \scalebox{0.95}{
                \begin{tabular}{llrrrrr}
                    \toprule
                    \bfseries Dataset & \bfseries Category & \bfseries ADA & \bfseries KNN & \bfseries LR & \bfseries RF & \bfseries SVM \\
                    \midrule
                    \midrule
                    \multirow{4}{*}{Cambridge} & Time
                    & 67.17(.04)
                    & 77.96(.07)
                    & 76.01(.07)
                    & 78.21(.05)
                    & \textbf{78.78}(.07)
                    \\
                    & Spectral
                    & 87.09(.04)
                    & 85.34(.05)
                    & 84.17(.06)
                    & \textbf{87.15}(.05)
                    & 84.84(.07)
                    \\
                    & Cepstral
                    & 83.84(.05)
                    & 85.56(.07)
                    & 83.27(.06)
                    & 87.82(.07)
                    & \textbf{87.15}(.06)
                    \\
                    & Tonal
                    & \textbf{84.74}(.09)
                    & 81.04(.05)
                    & 81.44(.04)
                    & 81.11(.07)
                    & 82.59(.07)
                    \\
                    \midrule
                    \multirow{4}{*}{Coswara-deep} & Time
                    & 55.65(.07)
                    & 62.34(.02)
                    & 54.21(.09)
                    & \textbf{64.65}(.05)
                    & 63.94(.07)
                    \\
                    & Spectral
                    & 65.77(.07)
                    & 68.18(.04)
                    & 72.03(.05)
                    & 71.76(.06)
                    & \textbf{74.46}(.06)
                    \\
                    & Cepstral
                    & 70.83(.06)
                    & 71.03(.03)
                    & 75.01(.05)
                    & \textbf{77.55}(.06)
                    & 75.62(.08)
                    \\
                    & Tonal
                    & 69.29(.06)
                    & 66.27(.02)
                    & 68.02(.03)
                    & 72.32(.06)
                    & \textbf{72.98}(.03)
                    \\
                    \midrule
                    \multirow{4}{*}{Coswara-shallow} & Time
                    & \textbf{61.63}(.04)
                    & 55.05(.06)
                    & 56.16(.09)
                    & 54.27(.07)
                    & 55.90(.09)
                    \\
                    & Spectral
                    & 66.69(.04)
                    & 61.02(.05)
                    & 69.85(.05)
                    & 69.15(.05)
                    & \textbf{72.32}(.04)
                    \\
                    & Cepstral
                    & 63.13(.09)
                    & 68.35(.04)
                    & 65.83(.03)
                    & \textbf{71.79}(.06)
                    & 70.62(.04)
                    \\
                    & Tonal
                    & 58.37(.08)
                    & 63.98(.05)
                    & 65.21(.08)
                    & 67.17(.08)
                    & \textbf{68.81}(.08)
                    \\
                    \bottomrule
                \end{tabular}
                }
            \end{table}

            BC signal domain results confirmed SVM and RF as the best performing models (\Cref{tab:featureCategoryResults}). Considering SVM's BC ROC-AUC across all datasets, we note that the 4 feature categories were broadly ranked in increasing predictive efficiency (Cambridge, Coswara-deep, Coswara-shallow): \emph{time domain} (79\%, 64\%, 56\%), \emph{tonal} (83\%, 73\%, 69\%), \emph{spectral} (85\%, 74\%, 72\%), and \emph{cepstral} (87\%, 76\%, 71\%). Spectral and cepstral categories achieved similarly high accuracies. Interestingly, the same ranking was prevalent for all 5 ML models, leading to the conclusion that the cepstral and spectral feature categories encode particularly informative COVID-19 data from breath and cough signals. A Repeated Measures ANOVA test \cite{iantovics2021black} confirms that the feature domains lead to statistically significant differences in ROC score for all three datasets ($p<0.02$). 
            
        \textbf{Individual features.} \enspace{}
            We start with the best-performing SVM classifier before broadening to include all models to identify general predictive efficiency patterns. The results forming the basis of our analysis are available in~\Cref{tab:svmFeatureEval}. A Repeated Measures ANOVA test \cite{iantovics2021black} verifies that the sample type leads to statistically significant differences in ROC score across all datasets ($p<0.05$). 
            
            \begin{table}[!t]
                \small
                \centering
                \caption{\emph{5-fold CV ROC-AUC as mean(std)}. The majority of features showed the most accurate results on the BreathCough (BC) vector. Feature categories were ranked in increasing accuracy: time domain, tonal, spectral, and cepstral.}\label{tab:svmFeatureEval}
                \begin{subtable}[t]{0.6\linewidth}
                    \centering
                    \caption{SVM, Cambridge data.}\label{tab:kddSvmFeatureCategories}
                    \scalebox{0.95}{
                        \begin{tabular}{llrrr}
                            \toprule
                            & & \multicolumn{1}{c}{B} & \multicolumn{1}{c}{C} & \multicolumn{1}{c}{BC} \\
                            \midrule
                            \midrule
                            All & All & 85.86(.07) & 85.80(.05) & 87.68(.06) \\
                            \midrule
                            \multirow{3}{*}{Time} & All & 72.77(.04) & 74.90(.08) & 78.78(.07) \\
                            & RMSE & 72.28(.05) & 76.45(.08) & 77.88(.08) \\
                            & ZCR & 64.59(.08) & 69.73(.06) & 71.40(.06) \\
                            \midrule
                            \multirow{7}{*}{Spectral} & All & 85.28(.06) & 84.03(.07) & 84.84(.07) \\
                            & S-BW & 69.24(.08) & 71.57(.04) & 75.4(.08) \\
                            & S-CENT & 73.45(.08) & 70.06(.08) & 78.07(.07) \\
                            & S-CONT & 86.14(.06) & 84.03(.08) & 85.98(.08) \\
                            & S-FLAT & 74.22(.07) & 75.44(.05) & 75.87(.06) \\
                            & S-FLUX & 79.70(.08) & 77.14(.06) & 82.08(.06) \\
                            & S-ROLL & 70.70(.07) & 67.22(.04) & 71.22(.06)  \\
                            \midrule
                            \multirow{4}{*}{Cepstral} & All & 86.25(.06) & 83.98(.06) & 87.15(.06) \\
                            & MFCC & 86.56(.04) & 83.25(.05) & 87.68(.04) \\
                            & MFCC-\(\Delta{}\) & 84.21(.04) & 79.67(.08) & 85.54(.08) \\
                            & MFCC-\(\Delta{}^2\) & 84.25(.09) & 78.29(.07) & 85.24(.09) \\
                            \midrule
                            \multirow{5}{*}{Tonal} & All & 79.69(.07) & 78.06(.07) & 82.59(.07) \\
                            & C-CQT & 76.29(.06) & 71.12(.09) & 77.30(.06) \\
                            & C-ENS & 77.56(.07) & 72.11(.07) & 83.50(.03) \\
                            & C-STFT & 77.57(.05) & 72.65(.03) & 77.78(.07) \\
                            & TN & 74.28(.04) & 70.85(.04) & 77.57(.05) \\
                            \bottomrule
                        \end{tabular}
                    }
                \end{subtable}
                    \hfill
                \begin{subtable}[t]{0.37\linewidth}
                    \centering
                    \caption{SVM, Coswara-deep data.}\label{tab:cosdeSvmFeatureCategories}
                    \scalebox{0.95}{
                        \begin{tabular}{rrr}
                            \toprule
                            \multicolumn{1}{c}{B} & \multicolumn{1}{c}{C} & \multicolumn{1}{c}{BC} \\
                            \midrule
                            \midrule
                            76.79(.04) & 70.85(.06) & 77.15(.05) \\
                            \midrule
                            61.80(.04) & 58.58(.06) & 63.94(.07) \\
                            55.89(.10) & 61.14(.07) & 61.81(.07) \\
                            64.68(.03) & 59.45(.13) & 64.60(.04) \\
                            \midrule
                            76.34(.05) & 66.74(.05) & 74.46(.06) \\
                            61.63(.07) & 63.51(.05) & 65.46(.04) \\
                            68.53(.06) & 59.91(.06) & 71.95(.05) \\
                            74.89(.05) & 63.42(.08) & 73.57(.09) \\
                            61.77(.08) & 59.86(.06) & 61.14(.03) \\
                            63.79(.06) & 62.76(.07) & 67.20(.04) \\
                            65.35(.05) & 63.16(.05) & 67.58(.08) \\
                            \midrule
                            74.57(.03) & 70.15(.09) & 75.62(.08) \\
                            74.24(.03) & 70.74(.01) & 75.38(.05) \\
                            64.85(.07) & 68.90(.05) & 68.99(.04) \\
                            66.65(.08) & 67.72(.06) & 70.72(.07) \\
                            \midrule
                            71.74(.05) & 64.06(.06) & 72.98(.03) \\
                            67.87(.04) & 62.78(.07) & 61.50(.05) \\
                            70.03(.07) & 65.14(.03) & 65.96(.05) \\
                            67.01(.05) & 61.80(.08) & 68.19(.10) \\
                            60.90(.04) & 62.84(.02) & 61.33(.03) \\
                            \bottomrule
                        \end{tabular}
                    }
                \end{subtable}
            \end{table}

            The majority of the 15 features significantly outperformed random guesses for COVID-19 classification across all datasets and sample types. The lowest accuracies were achieved by Coswara-shallow, matching previous findings. Similarities between Cambridge and Coswara-deep were underlined by sample types: BC achieved the highest mean ROC-AUC scores on average, whereas Coswara-shallow was split evenly between B and C. However, given all considered features, the Coswara-shallow dataset still showed its highest accuracy on BC samples since cepstral and tonal features were the most influential overall.  MFCC (cepstral), S-CONT (spectral), and C-ENS/C-STFT (tonal) were the highest-scoring features in their categories, whereas the time domain was more variable.

\begin{figure}[!t]
                \centering 
                \includegraphics[width=\textwidth]{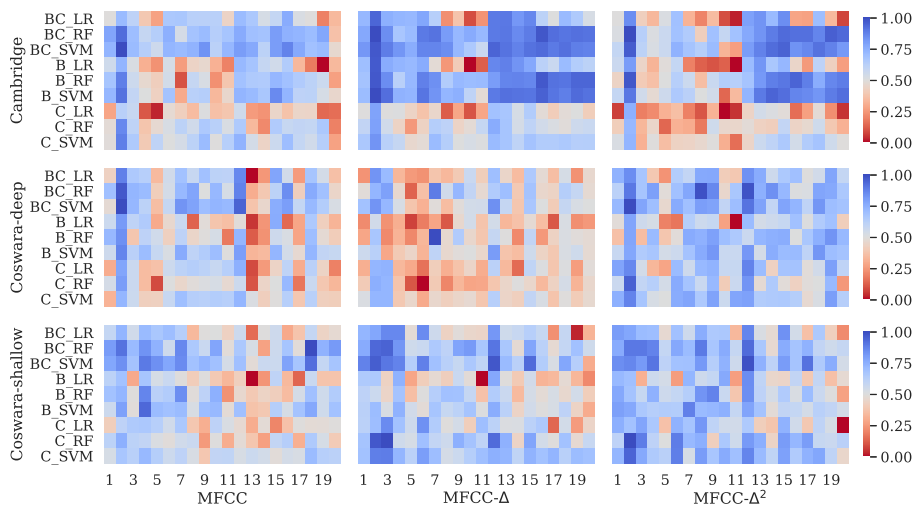}
                \caption{\emph{Normalised ROC-AUC of MFCC and derivatives for BreathCough (BC), Breath (B), and Cough (C) vectors.} Contrary to a common rule of thumb~\cite{brown2020exploring,jiang2002music,sharma2020coswara}, 13+ features provided significant discriminatory data, and showed that timbral information is especially relevant to COVID-19 classification.}\label{fig:mfccHeatmaps}
            \end{figure}

            Lastly, we note a surprising trend for MFCC.\@ A prevalent rule of thumb suggests 12--13 coefficients for audio classification~\cite{brown2020exploring,jiang2002music,sharma2020coswara}. However, \Cref{fig:mfccHeatmaps} shows that higher-order features provided discriminative information for COVID-19 on par with (Coswara-deep) or significantly outperforming (Cambridge) lower orders. This phenomenon was most noticeable in BC/B vectors and MFCC features. Since higher-order features contain information about details such as pitch and tone quality~\cite{MITROVIC201071}, we extrapolate that timber is highly relevant to COVID-19.

        \textbf{Discussion.} \enspace{}
            Our extensive analysis, comparison, and ranking of 15 features has found recurring patterns of predictive efficiency for COVID-19 audio classification across independent datasets. There was a distinct category ranking consistent across models, sample types, and datasets (increasing): time domain, tonal, spectral, and cepstral. Contrary to the intuitive expectation, some `complex' categories provided less discriminative information than `simpler' ones (e.g.\ tonal/spectral features). However, this is justified when considering that tonal features describe pitch and so are more suited to tasks with melodic content.

            The ranking underlines the significance of frequency-based features by elevating the spectral and cepstral categories describing timbral aspects and tone quality/colour. We have also shown that the common guideline to use only the first 13 MFCC features~\cite{brown2020exploring,jiang2002music,sharma2020coswara} was not applicable to COVID-19. Indeed, the higher-order (timbre) features' predictive efficiency provided significantly more discriminatory information, especially for the BC and B feature vectors.

            Taking a step back from the individual features, we note that the most prevailing pattern across all previous descriptions was that the concatenated BC feature vector outperformed the individual B and C vectors in most cases.

            Given our insights, we compare our results to the published baselines, summarised in~\Cref{tab:baselineAccuracyComparison}. The evaluated models were of similar type and complexity; The major difference was our introduction of new training features. We observe that our improved feature vectors significantly outperformed both the Cambridge and Coswara baseline accuracies by 10--17\%, validating our feature selection.

            \begin{table}[!t]
                \small
                \centering
                \caption{\emph{Comparison to dataset papers' 5-fold CV baseline results.} The most comparable configurations are shown (feature processing and classification model).}\label{tab:baselineAccuracyComparison}
                \begin{tabular}{clllrrr}
                    \toprule
                    \bfseries Origin & \bfseries Dataset & \bfseries Sample & \bfseries Model & \bfseries ROC-AUC & \bfseries Precision & \bfseries Recall \\
                    \midrule
                    \midrule
                    This paper & Cambridge & BC & SVM & \textbf{87.68}(.06) & 87.61(.07) & 81.39(.07) \\
~\cite{brown2020exploring} & Cambridge & BC & LR & 71.00(.08) & 69.00(.09) & 66.00(.14) \\
                    \midrule
                    This paper & Cos-deep & BC & SVM & \textbf{77.15}(.05) & 76.7(.05) & 53.09(.03) \\
~\cite{muguli2021dicova} & Cos-Unknown & C & RF & 67.45(---) & \multicolumn{1}{c}{---} & \multicolumn{1}{c}{---} \\
                    \bottomrule
                \end{tabular}
            \end{table}

\section{Related work}\label{sec:relatedWork}
    During in- and exhalation, air travelling through the respiratory tract undergoes turbulence and produces sounds. Consequently, any physical changes to the airways or lungs (e.g.\ caused by diseases such as COVID-19) also alter the produced respiratory sounds~\cite{rizal2015signal}. Even though listening and evaluating lung sounds manually is inherently subjective, medical professionals have long used this technique to non-invasively diagnose a wide variety of respiratory diseases~\cite{aykanat2017classification}.
    
    The popularisation of digital signal processing techniques and Machine Learning (ML) have made the automatic classification of respiratory sounds possible as a less subjective, low-cost, and patient-friendly (pre-)screening method. A literature review of existing implementations shows that ML can reliably pick up on subtle cues in audio signals for a variety of diseases.
    
    Smartwatches and wearable devices have made audio monitoring for healthcare purposes feasible. Nguyen et al.\ apply a dynamically activated respiratory event detection mechanism to detect cough and sneeze events non-intrusively~\cite{nguyen2018cover}. \cite{amrulloh2015cough} presents classifiers distinguishing between asthma and pneumonia in pediatric patients. Lastly, an image classification solution with comparable results is developed in~\cite{aykanat2017classification}, using spectrograms as the input.

    One of the first COVID-19 audio datasets containing breath and cough samples was presented in~\cite{brown2020exploring}. Using standard ML and audio processing techniques, the authors report 71\% ROC accuracy for COVID classification. ~\cite{muguli2021dicova} and~\cite{sharma2020coswara} consider further recording types such as vowel intonation and sequence counting, achieving 67\% and 66\% accuracy with ML models respectively.

\section{Conclusion and future work}\label{sec:conclusion}
    Our extensive comparative analysis of 15 audio features has provided significant insight into Machine Learning (ML) feature selection for COVID-19 respiratory audio classification and addressed the research questions laid out in~\Cref{subsec:researchQuestions}. Primarily, we identified the most informative feature characteristics and verified their ranking across two independent datasets. Since the two feature rankings showed considerable overlap, we conclude that the features' relative salience was likely inherent to the respiratory signals rather than the evaluated datasets.
    
    Throughout our analysis, a number of informative audio features were newly incorporated in the context of COVID-19 classification. In combination with our feature ranking, we achieved 88\% and 77\% accuracy on the Cambridge and Coswara datasets. Since the complexity of the signal processing and ML models is comparable to the baselines, the increase of up to 17\% and 10\% respectively was a consequence of our feature selection. Our established feature ranking could benefit future sound-based COVID-19 classification applications.
    
    This paper provides a starting point for the holistic evaluation of respiratory audio features for COVID-19 classification. Considerations that could be addressed in future work are a comprehensive strategy to regularise different sample lengths, and to identify the most informative audio features for complex architectures such as Deep Learning neural networks.
    
    Although sound-based COVID-19 detection was the primary purpose of this research, many other respiratory diseases and disorders could benefit from the development and improvement of automatic audio detection systems for diagnosis, treatment, and management. Therefore, the approach described in this paper could be generalised for the detection of other respiratory diseases.

\subsubsection{Acknowledgements} We would like to thank Chris Watkins for the stimulating discussions, and University of Cambridge for access to the COVID-19 sound dataset. This research is funded by University of Brighton's Connected Futures, Radical Futures' initiatives, and Santander's Global Challenges Research grant.

\bibliographystyle{splncs04}
\bibliography{bibliography}

\end{document}